\documentclass[10pt]{article}

\newcommand{\bb}{\begin{equation}}
\newcommand{\ee}{\end{equation}}
\newcommand{\bbb}{\begin{eqnarray}}
\newcommand{\eee}{\end{eqnarray}}

\begin{document}

\small{

\title{Comment on 
``On physical interpretation of the Poynting-Robertson effect"}
\author{R. Srikanth\thanks{e-mail for PostScript or
offprint requests of Srikanth (1999): srik@iiap.ernet.in} \\
Indian Institute of Astrophysics, Bangalore- 34, India.}
\date{}
\maketitle

\begin{abstract} 
In response to the comments of astro-ph/0006426 on Srikanth
(Icarus 1999), we wish to clarify, and thereby justify, the definition of
Poynting-Robertson drag adopted in the latter. We confirm
that dust absorption is a necessary condition while 
re-emission is neither necessary nor sufficient
for the inspiralling of a rest-isotropically emitting dust.
\end{abstract}

\section{Introduction}

Poynting-Robertson (P-R) effect causes small bodies
in circum-solar orbit, such as dust, assumed
to totally absorb intercepted radiation and re-emit isotropically in the 
bodies' rest-frame (``rest-isotropically"), to inspiral with
orbits of decreasing eccentricity (Harwit 1985; Burns 1979). 
The paradoxical, and hence interesting, element of the effect is that 
outward-directed raditation from a 
primary (the Sun or a star) can cause an otherwise stably orbiting dust
to infall.  Although the effect is
familiar enough through treatment in standard textbooks, a physical 
understanding
of its origin has been unclear (for a historical introduction, see Robertson 
1937; Kla{\v c}ka 1993; Srikanth 1999). 

More often than
not the drag has been attributed to the front-back asymmetry of the dust
re-emission as seen in the heliocentric frame. The
valid objection to this is that rest-symmetric emission should
not alter the intertial state of the center-of-mass, as is evident in 
the instantaneous rest-frame of the dust. In response to this,
the aberration of sunlight is invoked to explain the drag in the dust frame.
But to {\em this} we raise the objection that it is inconsistent to invoke 
two different causes (dust-emission asymmetry / sunlight aberration) in the
two different frames (heliocentric frame / instantaneous dust rest-frame).

The fact of a given cause engendering an observed effect is absolute, and
not relative to the choice of reference frame. 
The covariance of the force equations automatically enforces this:
the status of a given tensor as vanishing or non-vanishing is itself
absolute, even though its components transform covariantly. The
belief that the cause of the P-R effect is also frame-dependent is a popular
fundamantal misconception surrounding the effect. A subtlety to
be aware of is frame-dependent manifestations of a single cause (eg., 
the magnetic force between two current carrying wires transforms into an
electric in the rest frame of the valence electrons; this happens basically
because these two fields are part of a single Maxwell tensor field).
However, in the P-R effect, absorption and re-emission are in principle
distinct processes and transform independently of each other.
Strangely, this curious state of impasse doesn't seem to have attracted 
sufficient mainstream attention. 

In Srikanth (Icarus 1999), the P-R ``paradox" was resolved by carefully
distinguishing the contributions to the dust dynamics from the 
absorption and re-emission. We showed that
solar radiation absorption plays an explicit role
in the drag (i.e, azimuthal slow-down) of the dust, while re-emission, assumed
rest-isotropic, does not. The slow-down may be visualized
as a consequence of dust absorption and angular momentum conservation.

The present paper is intended to further clarify this result,
and in specific to 
respond to comments made in Kla{\v c}ka (astro-ph/0006426: hereafter 
Kla{\v c}ka). 
This is worth the effort, since history affirms that the paradoxical nature
of this effect can cause confusion.

\section{The equations of motion}

It is sufficient for our purpose to consider a
simple model of a spherical, completely absorbing dust, with the added
generalization that the absorption and re-emission parameters of the dust
are mutually distinct. 
The generally covariant equations governing the motion of the dust
can be written as:-
\bb
\label{mo0}
\frac{Dp^{\mu}}{D\tau} \equiv \frac{dm}{d\tau}u^{\mu} + 
m \frac{Du^{\mu}}{D\tau} = f_{ext}^{\mu},
\ee
where $p^{\mu} = mu^{\mu}$ is four-momentum of the dust, $\tau$ proper time,
$f_{ext}^{\mu}$ is the external four-force, and 
the operator $D/D\tau$ is the ``total" covariant derivative in the
General Relativistic sense and includes gravitational effects. Thus 
gravitation is not part of $f_{ext}^{\mu}$. The mass change term accounts for 
possible change in the dust's internal energy due to heating or cooling
(Srikanth 1999),

The primary radiation is considered as a plane-parallel beam flowing
radially outward, represented by the force
$f_{rad}^{\mu} =  \epsilon l^{\mu}$, where $l^{\mu}$ is a dimensionless
null-vector, with its spatial part being purely radial (Robertson 1937). 
The scalar $\epsilon$ $(> 0)$ is the rest-rate of absorption of solar radiation 
by the dust, with dimension momentum over time. 
We assume that the dust absorbs all the incident radiation
and re-emits  rest-isotropically.
The relativistic four-force associated with a rest-isotropic 
emission is the time-like four-vector $f_{emit}^{\mu} = - (\xi/c^{2}) u^{\mu}$, 
where $u^{\mu}$ is four-velocity, and $\xi$ $(> 0)$ is the rest-frame 
energy emission rate. Accordingly,  Eq. (\ref{mo0}) becomes:- 
\begin{eqnarray}
\label{mo01}
\frac{dm}{d\tau}u^{\mu} + 
m \frac{Du^{\mu}}{D\tau}& = &f_{rad}^{\mu} + f_{emit}^{\mu},\nonumber\\
&=& \epsilon l^{\mu} - (\xi/c^{2}) u^{\mu}.
\end{eqnarray}
This can be split into scalar and spacelike componants as follows.

Contracting Eq. (\ref{mo01}) by $u_{\mu}$, we get the scalar equation:
\begin{equation}
\label{mass}
c^{2}\frac{dm}{d\tau} = \epsilon l^{\alpha}u_{\alpha} - \xi.
\end{equation}
Substituting this back into Eq. (\ref{mo01}), we get the true equations of
motion:-
\begin{equation}
\label{motion}
m(\tau )\frac{Du^{\mu}}{D\tau} = \epsilon
\left(l^{\mu} -  \frac{l^{\alpha}u_{\alpha}}{c^{2}}u^{\mu}\right),
\end{equation}
where the bracketed term is just the part of 
radiation orthogonal to $u^{\mu}$.  The significance of this split is that
Eq. (\ref{mass}) describes the internal energy change of the dust but not
its motion, while Eq. (\ref{motion}) describes its motion. Both equations
are in general coupled because of the time-dependence of mass.
Implementing the metric for a weak
static gravitational field, and letting $v \ll c$, Eq. (\ref{motion}) can be 
shown to lead to the usual non-relativistic equations of motion for the
P-R effect (Srikanth 1999). 

\section{Identifying the P-R drag}

The first term on the right-hand side of Eq. (\ref{motion})
is the radiation pressure term. The second term, which
is formally the relativistic generalization of a friction force, 
is responsible for the azimuthal deceleration of the dust. 
Hence it is called the drag term. 
Switching it off, as it were, leads to a stably orbiting dust with 
the gravitational field modified by radiation pressure. To check that this
is the case, in Eq. (\ref{mo01}), we put $\epsilon = 0$, which leads to: 
\begin{eqnarray}
\label{mass0a}
c^{2}\frac{dm}{d\tau} &=& - \xi. \\
\label{mass0b}
m(\tau )\frac{Du^{\mu}}{D\tau} &=& 0.
\end{eqnarray}
Eq. (\ref{mass0b}) shows that in the absence of absorption, the dust moves
along a geodesic. The mass of the dust is time-dependent, but does not
affect its motion. Hence absorption is a necessary condition for 
dust slow-down, while re-emission is {\em not} a sufficient condition for
slow-down.

On the other hand, associating the four-momentum carried away from the dust,
which is the second term in the r.h.s of Eq. (\ref{mo01}),
with the drag, as Kla{\v c}ka recommends, is misleading. In Eq. (\ref{mass}),
putting $\xi = 0$ leads to: 
\begin{equation}
\label{mass1a}
c^{2}\frac{dm}{d\tau} = \epsilon l^{\alpha}u_{\alpha}, 
\end{equation}
while Eq. (\ref{motion}) remains unchanged. 
Thus, we still have the slow-down of the
dust in the absence of re-emission, though the trajectory is modified by
the mass change. This shows that re-emission is not a necessary condition
for slow-down, while absorption is a sufficient condition (assuming 
all the usual non-P-R aspects of problem as given). If $dm/d\tau \ne 0$ then
re-emission contributes implicitly via the time-dependent mass. However,
this is not a friction or drag-like contribution. Depending on whether 
the dust heats up
or cools down, this process renders the dust less or more susceptible to
drag.

In summary, re-emission is neither a necessary nor sufficient condition for
the drag, while absorption is necessary. The association of
re-emission with the drag is erroneous except 
in the case of an isothermal ($dm/d\tau = 0$) dust,
as seen via Eq. (\ref{mass}). Because
historically the P-R effect had been studied with an implicit assumption of
isothermality, this equality led to a general confusion of the 
relative roles played by these two processes in the drag, as discussed in the
Introduction. Indeed, the two-parameter model of the P-R effect that we
have adopted was motivated by this observation.
 
\section{Isothermality} 

As seen from its rest frame, an isothermal dust emits as much as it absorbs 
(Srikanth 1999). However, in any other frame, as seen from Eq. (\ref{mo01}), 
the excess of absorbed 
energy over re-emission is balanced by kinetic energy. Kla{\v c}ka argues that
this distinction between the reference frames is
obvious, and no more necessary to make explicit than to state
that the laws of reflection must be referred to the rest frame of the
reflecting surface.  

However, in fact, this analogy is not apt. Here, the 
rest-frame of the reflector is a natural choice for reference frame. The
simplest statement of the law of reflection also refers to this frame. But,
in the P-R effect, while the convenient choice of frame is the heliocentric
frame, the simplest possible satisfaction of isothermality (by the equality
of emitted to absorbed radiation) holds good in the dust rest-frame. This
has lead in some extant literature to the erroneous association of the
simplest version of isothermality with the heliocentric reference frame.
The purpose of mentioning this point in Srikanth (1999) was of pedagogical and
historical interest.

\section{Conclusions}

Some clarifications on an earlier
result reached by us in Srikanth (1999) are given. We find that in
the P-R effect 
dust absorption is a necessary condition for drag, whereas re-emission,
assumed to be rest-isotropic, is neither necessary nor sufficient. 
This result is independent of the reference frame used for the
description of the effect.

\section*{Acknowledgements}

I am thankful to Dr. S. P. K. Rajaguru and Dr. Gajendra Pandey.

}
\end{document}